\documentclass[reprint,aip,amsfonts,amsmath,floatfix]{revtex4-1}

\usepackage[USenglish]{babel} 
\usepackage[T1]{fontenc}
\usepackage[ansinew]{inputenc}

\usepackage{lmodern}

\usepackage{graphicx}
\usepackage{siunitx}
\usepackage{bm}
\usepackage{mathrsfs}
\DeclareMathAlphabet{\mathscrbf}{OMS}{mdugm}{b}{n}

\newcommand{\sub}{_\text}

\newcommand{\graphdir}{}

\newcommand{\etal}{\textit{et al.}}

\renewcommand{\vec}[1]{\boldsymbol{#1}}

\newcommand{\Rtens}{{\mathcal{R}}}

\newcommand{\gao}{$\beta\text{-Ga}_2\text{O}_3$}

\begin{document}

\title{Raman tensor elements of \gao}
\author{Christian Kranert}\email{christian.kranert@uni-leipzig.de}
\affiliation{Universit\"at Leipzig, Institut f\"ur Experimentelle Physik II, Abteilung Halbleiterphysik, Linn\'estra\ss e 5, 04103 Leipzig, Germany}
\affiliation{current address: Fraunhofer Technology Center for Semiconductor Materials THM, Am St.-Niclas-Schacht 13, 09599 Freiberg, Germany}
\author{Chris Sturm}
\author{R\"udiger Schmidt-Grund}
\author{Marius Grundmann}\email{grundmann@physik.uni-leipzig.de}
\affiliation{Universit\"at Leipzig, Institut f\"ur Experimentelle Physik II, Abteilung Halbleiterphysik, Linn\'estra\ss e 5, 04103 Leipzig, Germany}

\begin{abstract}

The Raman spectrum and particularly the Raman scattering intensities of monoclinic \gao\ are investigated by experiment and theory. The low symmetry of \gao\ results in a complex dependence of the Raman intensity for the individual phonon modes on the scattering geometry which is additionally affected by birefringence. We measured the Raman spectra in dependence on the polarization direction for backscattering on three crystallographic planes of \gao\ and modeled these dependencies using a modified Raman tensor formalism which takes birefringence into account. The spectral position of all 15 Raman-active phonon modes and the Raman tensor elements of 13 modes were determined and are compared to results from ab-initio calculations. 
\end{abstract}
\maketitle

\section{Introduction}

Gallium oxide in its stable $\beta$-modification is a semiconductor with a very wide bandgap of approximately \SI{4.8}{\electronvolt} \cite{GaO_bandgap1,GaO_bandgap2}. This makes this material an interesting candidate as active medium in deep UV optoelectronics. Further, this wide band gap suggests a larger theoretical breakdown field than for e.g. Si or SiC indicating a potential application of \gao\ in high power electronics \cite{GaO_FET_breakdown}.

For the use of the material in such applications, the knowledge of its fundamental properties is vital. These include the phonon energies as obtained by Raman spectroscopy which give access to sample properties like strain. In several experimental \cite{GaO_Raman_IR,GaO_Raman,GaO_Raman2} and theoretical \cite{GaO_Raman_IR,GaO_Raman,GaO_theory} works the energies of the Raman-active phonon modes have been reported. We briefly review these results and compare them to our own findings.

The main focus of the present paper is however the information that can be obtained from the Raman scattering intensities. Since \gao\ has a monoclinic crystal structure, its properties are strongly anisotropic. The investigation of Raman intensities provides access to the orientation of a particular sample via the selection rules and the dependence of the Raman scattering intensity for the individual phonon modes on the polarization relative to the crystal orientation. However, the well-known relation between scattering intensity $I$ and scattering geometry 
\begin{equation}
I\propto \left| e\sub{1}\Rtens e\sub{0}\right|^2
\label{eq:Raman_tensor_standard}
\end{equation}
with the polarizations of the incident and scattered light $e\sub{1}$ and $e\sub{0}$ at the point of the scattering event cannot be directly applied to experiments on anisotropic crystals. Owing to birefringence, the polarization of the radiation within the crystal, where the scattering event occurs, is in general elliptical and different from the incident and detected polarization $e\sub{i}$ and $e\sub{s}$ set by the experimental setup. Further, this effect is depth-dependent such that it was considered to be ``pointless'' \cite{principal_axes_only} to analyze the Raman intensities for polarizations which are not parallel to the principal axes of the dielectric indicatrix. Therefore, only Raman intensities for these polarization configurations were reported so far for \gao\ \cite{GaO_Raman_IR}. 

This experimental limitation causes severe limitations in the gain of knowledge. In application, it is of course desirable to calculate the scattering intensity for any orientation of the crystal in order to determine the orientation of a given sample. The restriction to polarization directions parallel to the principal axes further prevents the determination of the signs of the Raman tensor elements since only intensities are measured. Fortunately, for backscattering with sufficiently large scattering depth range, which is typically the case for bulk samples, the depth dependence vanishes and the scattering intensities can be described using a modified Raman tensor formalism \cite{RTFAC}. Here, we apply this formalism to model the Raman intensities for various scattering geometries and by that obtain the individual Raman tensor elements including their sign for most phonon modes of \gao .

\section{Methods}

We used three different cuts of \gao\ single crystals as samples: commercial, $(010)$- and $(\bar{2}01)$-oriented samples from Tamura Corporation and (100)-oriented samples from Leibniz-Institut f\"ur Kristallz\"uchtung (IKZ) Berlin. The crystals are free of twins which we verified by means of X-ray diffraction. Samples from both sources show identical Raman shifts within the margin of error. The same is true for a comparison between unintentionally doped and Sn-doped samples from Tamura. Thus, in the following, these specifics of the samples are neglected.

Raman scattering was excited using a diode-pumped solid state laser emitting at $\lambda = \SI{532}{\nano\meter}$. The incident light was focused on the sample by a microscope objective with magnification of $50\times$ and a numerical aperture of $NA=0.42$; the scattered light was collected by the same objective (backscattering geometry). We neglected the influence of oblique rays due to the focusing aperture. The sample was placed with its polished surface perpendicular to the direction of light propagation. An achromatic $\lambda/2$ waveplate was placed between beam splitter and objective and rotated by an angle of $\varphi/2$ in order to rotate the polarization of both the incident and detected radiation by an angle of $\varphi$ relative to the sample. The Glan-Thompson polarizer used as analyzer was kept fix. Another $\lambda/2$ waveplate was applied in front the beam splitter either at \SI{0}{\degree} or \SI{45}{\degree} to select between parallel polarization or cross polarization. The spectrum was recorded using a double spectrometer with $2\times\SI{1}{\meter}$ focal length, gratings with 2400 lines per mm and equipped with a liquid nitrogen-cooled charged-coupled device camera with a pixel pitch of \SI{13.5}{\per\centi\meter}. The slit width was set to \SI{100}{\micro\meter} yielding a spectral resolution of \SI{0.45}{\per\centi\meter}. All measurements were carried out at room temperature. The dielectric tensor as function of the wavelength and particular at the excitation wavelength was determined by means of spectroscopic ellipsometry published elsewhere \cite{ga2o3_df}.

Ab-initio calculations of the phonon energies and Raman intensities were carried out using the CRYSTAL14 code. We used the basis set of Pandey \etal\ \cite{basisset_Ga} for gallium and of Valenzano \etal\ \cite{basisset_O} (slightly modified) for oxygen. The \mbox{B3LYP} hybrid functional was used as it is known to yield good agreement to experimental results for vibrational properties \cite{basisset_O,B3LYP_comparison}. We could also verify this finding based on comparative calculations using other Hamiltonians (\mbox{B3PW}, \mbox{PBE0}, \mbox{HSE06}). Pack-Monckhorst and Gilat shrinking factors of 8 were used which corresponds to 150 $k$-points in the irreducible Brillouin zone. The truncation criteria for the Coulomb and exchange infinite sums are defined in the CRYSTAL14 code by five tolerances set to 8, 8, 8, 8, and 16 for our calculations. The tolerance for the energy convergence was set to $10^{-11}$ Hartree. The lattice parameter optimization with these parameters yields $a_0=\SI{12.336}{\angstrom}$, $b_0=\SI{3.078}{\angstrom}$, $c_0=\SI{5.864}{\angstrom}$ and $\beta=\SI{103.89}{\degree}$. This slight overestimation of the lattice parameters with respect to experimental values \cite{algao_raman_kranert,GaO_lattice_parameter,ingao_raman_kranert} is expected for the \mbox{B3LYP} hybrid functional \cite{b3lyp_metal,basisset_O,B3LYP_comparison}.

\section{Results and Discussion}

\subsection{Phonon energies}

The stable form of \gao\ under ambient conditions has a monoclinic symmetry which belongs to the space group $C2/m$ in international and $C_{2h}^3$ in Sch\"onflies notation. The [010]-direction is perpendicular to [100] and [001] which confine an angle $\beta=\SI{103.7}{\degree}$. Because the [010]-direction is perpendicular to the other two crystal axes lying in the (010)-plane, we choose $z \| [010]$ in the following. The assignment of the in-plane coordinates is discussed below and depicted in Fig. \ref{fig:orientation}.

The unit cell of \gao\ consists of 10 atoms which results in 30 phonon modes of which 27 are optical modes. At the $\Gamma$-point, these belong to the irreducible representation \cite{GaO_Raman_IR}
\begin{equation}
\Gamma^{opt} = 10 A_g + 5 B_g + 4 A_u + 8 B_u \, .
\label{eq:irred_rep}
\end{equation}
The modes with $A_g$ and $B_g$ symmetry are Raman-active, while those with odd parity (index $u$) are infrared active. Under non-resonant conditions, the Raman tensors for the Raman-active phonon modes have the form \cite{Raman_Hayes_Loudon}
\begin{equation}
A_g:\, \Rtens=\begin{pmatrix}	
a & d & 0 \\
d & b & 0 \\
0 & 0 & c
\end{pmatrix};\quad B_g:\, \Rtens=\begin{pmatrix}	
0 & 0 & e \\
0 & 0 & f \\
e & f & 0
\end{pmatrix} \, 
\label{eq:tensors}
\end{equation}
with real tensor elements.

\begin{figure}%
\includegraphics[width=\columnwidth]{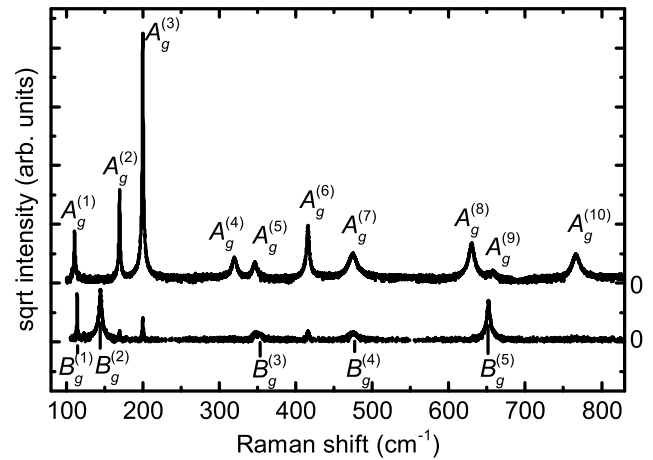}%
\caption{Experimental Raman spectra of two \gao\ single crystals. Top curve: (010)-oriented crystal, scattering geometry $z(yy)z$, bottom curve: $(\bar{2}01)$-oriented crystal, scattering geometry $z'(x'y')z'$, where $y\| [100]$, $x' \| z \| [010]$, $y'\|[102]$, and $z'\perp (\bar{2}01)$.}%
\label{fig:spectra}%
\end{figure}

The selection rules induced by the form of the Raman tensors allow to distinguish between the two types of vibrational symmetry as it is depicted by the experimental Raman spectra shown in Fig. \ref{fig:spectra}. For excitation on the (010)-plane, the $B_g$ modes are forbidden and the $A_g$ modes are allowed for all polarization configurations. For an excitation on a surface perpendicular to that both types of modes can be allowed, depending on the polarization configuration. In the cross polarized configuration shown in the bottom curve of Fig. \ref{fig:spectra}, only the $B_g$ modes are allowed. The not perfect extinction of the Raman lines due to the $A_g$ modes results from polarization leakage.
The extinction ratio is approximately 1:50.

The spectral positions of the individual Raman modes were obtained by modeling their spectral line shape using Lorentzian functions. These experimental results are summarized in Tab. \ref{tab:positions} and compared to our results from ab-initio calculations as well as to values from the literature. A very good agreement between theory and experiment can be observed. The comparison between the experimental results only show a good agreement between our results and those of Refs. \citenum{GaO_Raman_IR,GaO_Raman} with only small deviations for some phonon modes. We obtain identical results within a margin of error of \SI{0.2}{\per\centi\meter} when using an alternative excitation wavelength ($\lambda=\SI{325}{\nano\meter}$) or sintered powder samples which substantiates the trust in our results. 

\begin{table*}%
\begin{tabular}{llllllllll}
\hline\hline
phonon mode & \multicolumn{4}{c}{Experiment} & & \multicolumn{4}{c}{Theory} \\\cline{2-5}\cline{7-10}
 & this work & Ref. \citenum{GaO_Raman_IR} & Ref. \citenum{GaO_Raman} & Ref. \citenum{GaO_Raman2} & & this work & Ref. \citenum{GaO_Raman_IR} & Ref. \citenum{GaO_Raman} & Ref. \citenum{GaO_theory}\\ \hline
$A_g^{(1)}$  & 111.0 & 111 & 110.2 & 112 & & 113.5 & 113 & 104 & 104.7\\
$B_g^{(1)}$  & 114.8 & 114 & 113.6 & 115 & & 118.6 & 114 & 113 & 112.1\\
$B_g^{(2)}$  & 144.8 & 147 & 144.7 & 149 & & 145.6 & 152 & 149 & 141.3\\
$A_g^{(2)}$  & 169.9 & 169 & 169.2 & 173 & & 176.4 & 166 & 165 & 163.5\\
$A_g^{(3)}$  & 200.2 & 199 & 200.4 & 205 & & 199.1 & 195 & 205 & 202.3\\
$A_g^{(4)}$  & 320.0 & 318 & 318.6 & 322 & & 318.5 & 308 & 317 & 315.8\\
$A_g^{(5)}$  & 346.6 & 346 & 346.4 & 350 & & 342.5 & 353 & 346 & 339.7\\
$B_g^{(3)}$  & 353.2 & 353 & n.o.  & 355 & & 359.2 & 360 & 356 & 348.3\\
$A_g^{(6)}$  & 416.2 & 415 & 415.7 & 421 & & 432.0 & 406 & 418 & 420.2\\
$A_g^{(7)}$  & 474.9 & 475 & n.o.  & 479 & & 472.8 & 468 & 467 & 459.4\\
$B_g^{(4)}$  & 474.9 & 475 & 473.5 & 480 & & 486.1 & 474 & 474 & 472.8\\
$A_g^{(8)}$  & 630.0 & 628 & 628.7* & 635 & & 624.4 & 628 & 600 & 607.1\\
$B_g^{(5)}$  & 652.3 & 651 & 652.5* & 659 & & 653.9 & 644 & 626 & 627.1\\
$A_g^{(9)}$  & 658.3 & 657 & n.o.*  & 663 & & 655.8 & 654 & 637 & 656.1\\
$A_g^{(10)}$ & 766.7 & 763 & 763.9 & 772 & & 767.0 & 760 & 732 & 757.7\\
\hline\hline
\end{tabular}
\caption{Spectral position of the Raman peaks of the phonon modes of \gao , given in \SI{}{\per\centi\meter}. A more likely assignment for two peaks from Ref. \citenum{GaO_Raman} indicated by ``*'' was chosen to enhance the comparability, ``n.o.'' denotes modes which were not observed.}
\label{tab:positions}
\end{table*}

The comparison of the theoretical results shows a better agreement for our calculations and those of Dohy \etal \cite{GaO_Raman_IR} than obtained using the VASP \cite{GaO_Raman} and abinit code \cite{GaO_theory}, which both use plane-wave basis sets as opposed to the Gaussian-type orbital basis sets used by CRYSTAL14. The agreement for the calculations of Dohy \etal\ \cite{GaO_Raman_IR} is similarly good as that for our calculations, which is particularly remarkable regarding the very limited computing power available at the time of their work.

\subsection{Raman tensor elements}

\subsubsection{Theoretical background}

Raman scattering intensities were modeled using the formalism introduced by us in a preceding publication \cite{RTFAC}. In its general form, the Raman intensity in dependence on the polarization for normal-incidence backscattering on a certain surface is given by
\begin{equation}
I\propto | \vec{e}_\text{s} S J(z') T^\top R \Rtens R^{-1} T J(z') S^{-1}\vec{e}_\text{i} |^2 \, .
\label{eq:Raman_intensity_general}
\end{equation}
\noindent Here, $\vec{e}\sub{i}$ and $\vec{e}\sub{s}$ are the normalized polarization vectors of the incident and detected field, respectively. $S$ is the rotational matrix transforming the in-plane coordinates of the laboratory system into the in-plane coordinate system ($x'$, $y'$) spanned by the fast and the slow axis of the crystal. Further, $J(z')$ is the Jones matrix for light propagating along the surface normal $z'$. The transformation matrix 
\begin{equation}
T=\begin{pmatrix}
	1 & 0\\
	0 & \frac{\epsilon_{z'z'}}{\sqrt{\epsilon_{y'z'}^2+\epsilon_{z'z'}^2}}\\
	0 & \frac{\epsilon_{y'z'}}{\sqrt{\epsilon_{y'z'}^2+\epsilon_{z'z'}^2}}\\
\end{pmatrix}
\label{eq:transformation_matrix}
\end{equation}
\noindent transforms the external polarization, which is pinned to the $x'$-$y'$-plane, to the allowed internal polarizations, which may also exhibit an out-of-plane ($z'$) component if none of the principal axes is parallel to $z'$. The rotational matrix $R$ rotates the Raman tensor $\Rtens$, which is defined in the system of principal axes of the indicatrix $x$, $y$ and $z$, into the coordinate system $x'$, $y'$, $z'$ determined by the sample orientation.

Birefringence further causes different reflection coefficients at the surface. For normal incidence and vanishing absorption, it is described by the well-known equation
\begin{equation}
r_\text{fast,slow}=\frac{n_\text{fast,slow}-1}{n_\text{fast,slow}+1}
\label{eq:reflection}
\end{equation}
for the fast and slow axis, respectively. In order to properly model our experimental intensities, we took this into account by inserting the diagonal matrix $\rho = \mathrm{diag}(r_{x'x'}/r_{y'y'}, 1, 1)$ between $S$ and $J$. Since no absolute intensities were measured, only the relative reflectivity was considered. Equation \eqref{eq:Raman_intensity_general} then reads
\begin{equation}
I\propto | \vec{e}_\text{s} S \rho J(z') T^\top R \Rtens R^{-1} T J(z') \rho S^{-1} \vec{e}_\text{i} |^2 \, .
\label{eq:Raman_intensity_refl}
\end{equation}

Regarding the form of equation \eqref{eq:Raman_intensity_refl}, $\Rtens_\text{eff}(z') = \rho J(z') T^\top R \Rtens R^{-1} T J(z') \rho $ acts as a two-dimensional, depth-dependent effective Raman tensor. We have shown that for integration over a sufficient depth range, the depth dependence vanishes and the scattering intensity can be expressed as\cite{RTFAC}
\begin{equation}
I\propto | \vec{e}_\text{s} \Rtens_0 \vec{e}_\text{i} |^2 + | \vec{e}_\text{s} \Rtens_1 \vec{e}_\text{i} |^2 +| \vec{e}_\text{s} \Rtens_2 \vec{e}_\text{i} |^2
\label{eq:Raman_intensity_convergence}
\end{equation}
with the three components
\begin{subequations}
\begin{eqnarray}
\Rtens_0 &=& \begin{pmatrix}
	\Rtens_{\mathrm{eff},x'x'} & 0 \\
	 0 & 0
\end{pmatrix}\, , \label{eq:R0} \\ 
\Rtens_1 &=& \begin{pmatrix}
	0 & \Rtens_{\mathrm{eff},x'y'} \\
	 \Rtens_{\mathrm{eff},y'x'} & 0
\end{pmatrix}\, , \label{eq:R1} \\
\Rtens_2 &=& \begin{pmatrix}
	0 & 0 \\
	 0 & \Rtens_{\mathrm{eff},y'y'}
\end{pmatrix}\, . \label{eq:R2} 
\end{eqnarray}
\end{subequations}
\noindent Our previous calculations show that this approximation can be applied for typical experimental conditions for investigating bulk material. \cite{RTFAC} We verified the validity of this approximation for the present studies by measuring the actual depth intensity profile of our Raman setup and taking into account the birefringence for the crystal cuts of \gao\ measured by us (not shown). We found the approximation to be valid for all cuts under investigation.

The scattering intensity further depends on the phonon frequency $\omega\sub{P}$ and incident photon frequency $\omega\sub{i}$, i.e. explicitly $I\propto C(\omega) |\vec{e}\sub{s}\Rtens_\text{eff}\vec{e}\sub{i}|^2$ with \cite{Raman_Hayes_Loudon}
\begin{equation}
C(\omega\sub{P}) = \frac{\omega\sub{i}(\omega\sub{i}-\omega\sub{P})^3}{\omega\sub{P}\left[1-\exp\left(-\frac{\hbar\omega\sub{P}}{k\sub{B}T}\right)\right]} \, .
\label{eq:prefactor}
\end{equation}
In order to obtain Raman tensor elements which can be compared between the individual phonon modes, this prefactor must be taken into account.

\subsubsection{Determination of the Raman tensor}

\begin{figure}%
\includegraphics[width=6cm]{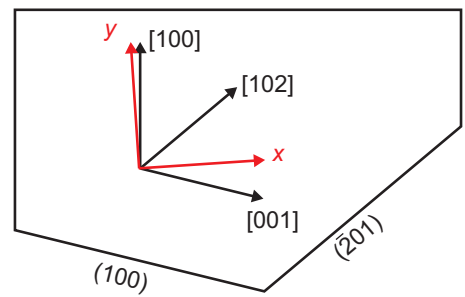}%
\caption{Relation between the crystallographic axes and the coordinate system used here. The [010]-direction points out of the drawing plane, the depicted directions are in the [010]-plane. The crystal cuts yielding (100)- and $(\bar{2}01)$-orientation are shown on the bottom. The angle between $y$ and [100] is \SI{3.8}{\degree}, that between $x$ and [001] is $\theta_{(100)}=\SI{-17.6}{\degree}$ and that between $x$ and [102] is $\theta_{(\bar{2}01)}=\SI{36.2}{\degree}$.}%
\label{fig:orientation}%
\end{figure}

Within the static approximation, i.e. for off-resonant excitation, the Raman tensor represents a modulation of the dielectric tensor. Therefore, it is defined in the coordinate system of the principal axes of the dielectric indicatrix which need to be known in order to analyze the Raman tensor. For the monoclinic system of \gao , one of these axes is parallel to the crystallographic [010]-direction. In order to determine the other two principal axes of the indicatrix, we used the dielectric tensor at the excitation wavelength derived from ellipsometry measurements \cite{ga2o3_df}
\begin{equation}
\epsilon (\lambda=\SI{532}{\nano\meter}) = \begin{pmatrix}
	3.669 & 0 & 0.0070\\
	0 & 3.809 & 0\\
	0.0070 & 0 & 3.774
\end{pmatrix} \, 
\label{eq:df532}
\end{equation}
given for a \emph{crystal} coordinate system with $x \| [100]$, $y \| [010]$ and $z$ chosen accordingly as used in Ref. \citenum{ga2o3_df}. From the eigenvectors of this tensor, one finds that the principal axes are tilted by an angle of \SI{3.8}{\degree} relative to the coordinate system defined above. In the following, the \emph{dielectric} coordinate system as indicated in red in Fig. \ref{fig:orientation} is used as coordinate system for which the Raman tensors take the form \eqref{eq:tensors}. Its axes are parallel to the principal axes of the dielectric indicatrix and the $z$-axis points in the [010]-direction.

\begin{figure*}[t]%
\includegraphics[width=14.5cm]{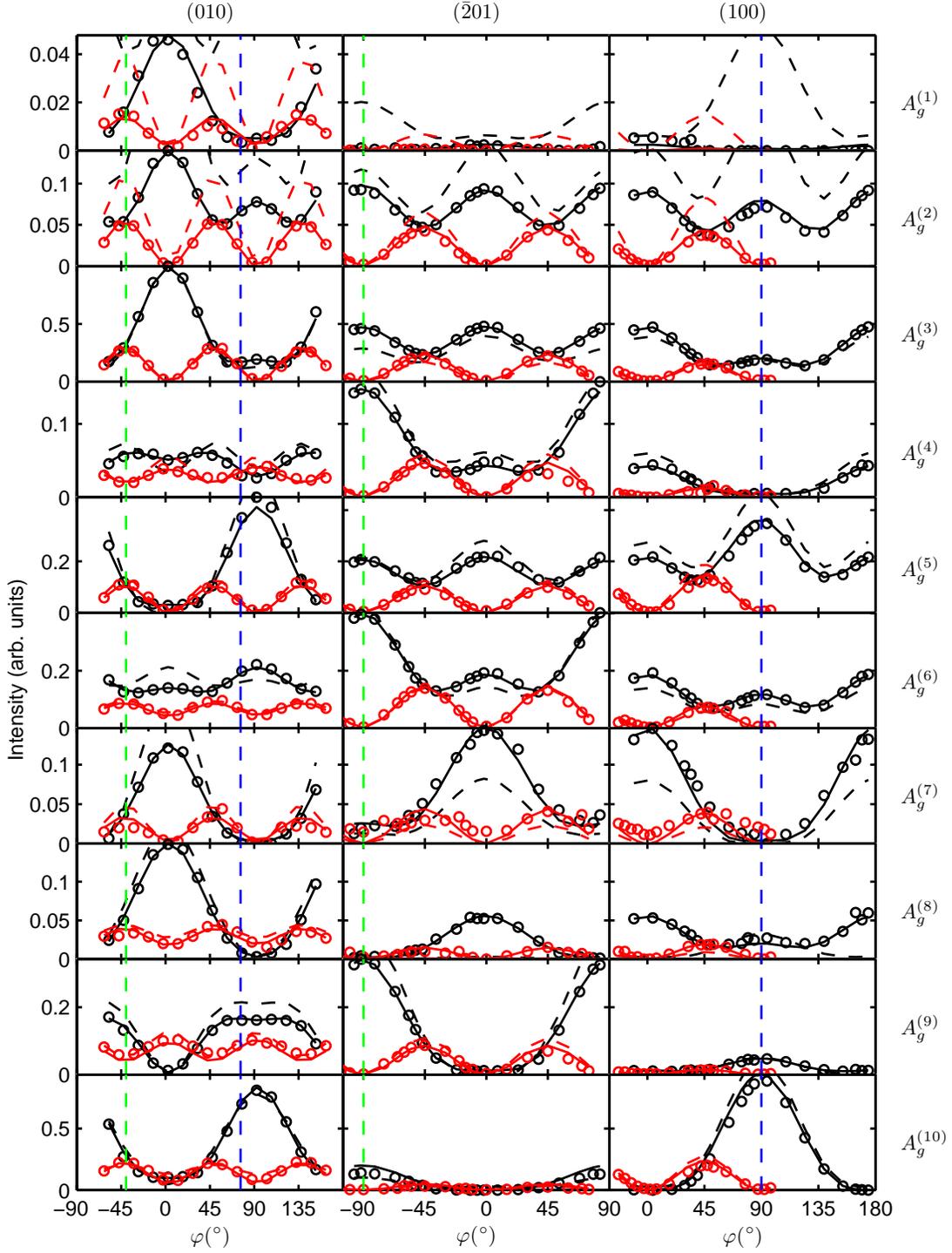}%
\caption{Experimental Raman scattering intensities (circles), model fits (solid lines) and modelled intensities from ab-intio-calculated tensor elements (dashed lines) for the phonon modes with $A_g$-symmetry of \gao\ in dependence on the direction of polarization $\varphi$. The orientation of the excited surface is indicated on top of the spectra and the phonon modes right to the spectra. An intensity of 1 means the overall maximum intensity which is observed for $A_g^{(3)}$. The intensity range is the same for all plots of a single phonon mode. For the $(010)$-orientation, $\varphi=\SI{0}{\degree}$ is set such that it coincides with the $[100]$-direction, for the other two orientations $\varphi=\SI{0}{\degree}$ corresponds to the $[010]$-direction. The dashed green and blue lines indicate the [102]- and [001]-direction, respectively.}%
\label{fig:Ag_fits}%
\end{figure*}

The three different crystal cuts investigated here are indicated in Fig. \ref{fig:orientation}.  We introduce a \emph{sample} coordinate system, indicated by a prime, such that the surface normal is parallel to $z'$. The in-plane coordinates are determined by the orientation of the slow and the fast axis and the transformation between sample and dielectric coordinate system is given by the rotation matrix $R$ as in equation \eqref{eq:Raman_intensity_refl}. For the (010)-plane, $z=z'$ such that no transformation is necessary and sample and dielectric coordinate system are identical. For the other two crystal cuts, one of the allowed polarization directions is parallel to the $z$-axis. With $z'$ being the direction of light propagation and $x'$ set parallel to $[010]$, $y'$ is obtained by rotating $x$ by an angle $\theta$ around $z$. Since the propagation direction is not parallel to a principal axis, the polarization component perpendicular to that is slightly tilted against the surface requiring to use the elements of the dielectric tensor to determine the matrix $T$ in equation \eqref{eq:Raman_intensity_refl}. Further, this polarization direction is not parallel to any of the principal axes. Thus, the scattering intensity for this direction results from a linear combination of Raman tensor elements, allowing to access their sign. Since no sample with the $[010]$-direction tilted against the surface was available for our research, this was only possible for the tensor elements $a$, $b$ and $d$. Further, the differences in reflectivity caused by birefringence were taken into account using the reflectivity ratio $r\sub{slow}/r\sub{fast}$ between the slow and the fast axis for the amplitude of the radiation. The parameters obtained from the dielectric tensor for the three surfaces required for modeling are summarized in Tab. \ref{tab:cuts}. As mentioned above, equation \eqref{eq:Raman_intensity_convergence} is expected to approximately hold for all investigated orientations. This is in agreement to the experiment judging from the line shapes in Fig. \ref{fig:Ag_fits}.

\setlength\tabcolsep{.15cm}
\begin{table}[tbp]%
\caption{Parameters for the crystal cuts under investigation. The slow axis (higher reflectance) is the $x$-axis for excitation on the $(010)$-facet and the $z$-axis for the other two facets.}
\begin{tabular}{l r r r r r}
\hline\hline
facet & $\theta$ & $\epsilon_{y'z'}$ & $\epsilon_{z'z'}$ & $\Delta n$ & $r\sub{slow}/r\sub{fast}$\\ \hline
$(010)$ & \SI{0}{\degree} & 0 & 3.809 & 0.027 & 1.0204 \\
$(\bar{2}01)$ & \SI{36.2}{\degree} & $-0.050$ & 3.705 & 0.018 & 1.0134 \\
$(100)$ & \SI{-17.6}{\degree} & $0.031$ & 3.678 & 0.011 & 1.0082 \\ \hline\hline
\end{tabular}
\label{tab:cuts}
\end{table}

\begin{figure}[tbp]%
\includegraphics[width=\columnwidth]{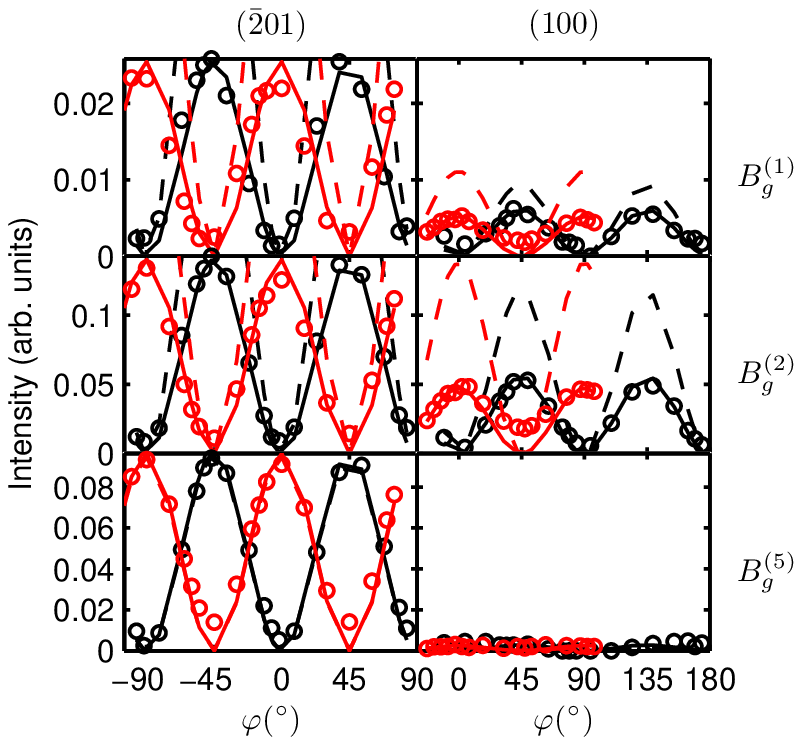}%
\caption{Same as Fig. \ref{fig:Ag_fits} for the phonon modes with $B_g$ symmetry as indicated. The intensity of $1$ is defined as in Fig. \ref{fig:Ag_fits}.}%
\label{fig:Bg_fits}%
\end{figure}

We carried out Raman measurements in backscattering geometry on the three mentioned surfaces with the polarization direction rotated by the angle $\varphi$ relative to the crystal. The intensities for the individual phonon modes were obtained as the area of Lorentzian functions used to model the line shape of the respective Raman peaks. Plotting these intensities over the angle $\varphi$ representing the direction of the polarization relative to the crystal yields graphs as in Figs. \ref{fig:Ag_fits} and \ref{fig:Bg_fits}. In order to assure comparability, the intensities were normalized to the maximum intensity of the prominent peak of the $A_g^{(3)}$ phonon mode at \SI{200}{\per\centi\meter}.



We determined the polarization-dependent Raman intensities for all ten phonon modes with $A_g$ symmetry and successfully modeled these dependencies with great agreement using the formalism for anisotropic crystals \cite{RTFAC} as depicted in Fig. \ref{fig:Ag_fits}. Only the four Raman tensor elements $a$, $b$, $c$ and $d$ were used as free parameter to fit the six polarization dependencies for each phonon mode. When necessary, the intensities for a measurement set were adjusted using an additional scale factor which is constant for all phonon modes. Similarly, a possible misalignment of the sample was adjusted using a constant angular offset for all modes. The required parameters resulting from the dielectric tensor were taken from the ellipsometry results as listed in Tab. \ref{tab:cuts}. 

The effect of birefringence can be clearly seen in the plots for the $A_g$ modes in Fig. \ref{fig:Ag_fits} by two effects: First, without mode conversion due to birefringence, the intensity for excitation on the $(\bar{2}01)$- and $(100)$-planes is expected to either vary between a maximum and minimum at $\phi=\SI{0}{\degree}$ and $\phi=\SI{90}{\degree}$, respectively, or to have an intensity of 0 in between. Second, the intensity for the polarization parallel to the [102]-direction is different for excitation on the $(\bar{2}01)$- and $(010)$-planes. The same is true for the polarization parallel to the [001]-direction on the $(100)$- and $(010)$-planes (see dashed lines in Fig. \ref{fig:Ag_fits}). This effect cannot be expected from \eqref{eq:Raman_tensor_standard}, but results from the fact that light with polarizations in these directions is split into two waves propagating in the crystal with different velocities, while this does not occur for the other two planes. In the latter case, additionally the tilting of the polarization relative to the surface must be considered, which is however a much smaller correction.

We modeled the observed intensity dependencies for each phonon mode individually, i.e. neglecting the phonon energy-dependent prefactor \eqref{eq:prefactor}. The Raman tensor elements were then obtained from the fitting parameters by dividing by $\sqrt{C(\omega\sub{P})}$. They are summarized in Tab. \ref{tab:tensor_elements_Ag}. They are normalized to the largest value which is the $a$ tensor element of the $A_g^{(10)}$ mode. Since the sign of the Raman tensor element $c$ could not be determined, these are all given as positive. As discussed below, it is reasonable to adopt the signs from the computed results. The signs of the other tensor elements are given with respect to $a$, i.e. $a$ is set as positive for all phonon modes. Please note that the Raman tensor elements enter quadratically in the scattering intensity.

\setlength\tabcolsep{.1cm}
\begin{table}%
\caption{Raman tensor elements obtained from modeling the polarization dependence of the experimental Raman scattering intensity and from theoretical calculations, normalized to a value of $1000$ for the highest Raman polarizability. Experimental values for $c$ are given as positive values, but may as well be negative from experimental results.}
\label{tab:tensor_elements_Ag}
\begin{tabular}{l rrrr c rrrr}
\hline\hline
 & \multicolumn{4}{c}{Experiment} & & \multicolumn{4}{c}{Theory}\\\cline{2-5}\cline{6-10}
 & \multicolumn{1}{c}{$a$} & \multicolumn{1}{c}{$b$} & \multicolumn{1}{c}{$c$} & \multicolumn{1}{c}{$d$} & & \multicolumn{1}{c}{$a$} & \multicolumn{1}{c}{$b$} & \multicolumn{1}{c}{$c$} & \multicolumn{1}{c}{$d$} \\
\hline
$A_g^{(1)}$ & 18 & $-59$   &  13 & 13    & &   79 &  $-70$ &    21 &     9 \\
$A_g^{(2)}$ & 104 &  144   & 117 & $-1$  & &  142 &   214  &   150 &  $-34$\\
$A_g^{(3)}$ & 187 &  $443$ & 396 & 15    & &  154 &  $431$ &   272 &  $-16$\\
$A_g^{(4)}$ & 103 &  $144$ & 132 & 125   & &  124 &  $113$ &   154 &  $146$\\
$A_g^{(5)}$ & 417 &  $120$ & 315 & $-6$  & &  479 &   $12$ &   349 &  $-18$\\
$A_g^{(6)}$ & 341 &  $288$ & 333 & 158   & &  320 &  $358$ &   293 &   146 \\
$A_g^{(7)}$ & 46  &  $-298$& 322 &$-51$  & &   31 & $-369$ & $-241$&     9 \\
$A_g^{(8)}$ & 52  &  $390$ & 238 & $-135$& &   55 &  $414$ &    53 & $-164$\\
$A_g^{(9)}$ & 401 &  $80$  & 115 & 321   & &  468 &   $61$ &    21 &   364 \\
$A_g^{(10)}$ &1000&  $356$ &   0 & $-270$& & 1000 &  $248$ & $-191$& $-409$\\
\hline\hline
\end{tabular}
\end{table}

Only three out of the five phonon modes with $B_g$ symmetry are shown in Fig. \ref{fig:Bg_fits}. The other two modes are too weak and for most scattering geometries superimposed by the spectrally close $A_g^{(5)}$ and $A_g^{(7)}$ modes. Further, modes with this symmetry cannot be observed for excitation on the (010)-plane owing to the selection rules. Therefore, they are only shown for the other two orientations in Fig. \ref{fig:Bg_fits}. From only these two orientations, the signs of the Raman tensor elements cannot be unambiguously determined, i.e. the polarization dependencies could be equivalently modeled using different Raman tensor elements with opposite signs. Again, the agreement between experiment and theory indicates that the alternative of experimental values given in the table is likely to be the correct one.

\begin{table}%
\caption{Raman tensor elements obtained from modeling the polarization dependence of the experimental Raman scattering intensity and from theoretical calculations, normalized to a value of 1000 for the tensor element $a$ of the $A_g^{(10)}$ mode. Elements marked with ``n.d.'' could not be determined.}
\label{tab:tensor_elements_Bg}
\begin{tabular}{l rr c rr}
\hline\hline
 & \multicolumn{2}{c}{Experiment} & & \multicolumn{2}{c}{Theory}\\ \cline{2-3} \cline{5-6}
 & \multicolumn{1}{c}{$e$} & \multicolumn{1}{c}{$f$} & & \multicolumn{1}{c}{$e$} & \multicolumn{1}{c}{$f$}\\
\hline
$B_g^{(1)}$ & 32 & 31 & & 46 & 56\\
$B_g^{(2)}$ & 106 & 70 & & 148 & 88\\
$B_g^{(3)}$ & n.d. & n.d. & & 238 & $-92$\\
$B_g^{(4)}$ & n.d. & n.d. & & 12 & $-291$\\
$B_g^{(5)}$ & 162 & 326 & & 147 & 335\\
\hline\hline
\end{tabular}
\end{table}

Theoretical values for the Raman tensor elements were obtained using the Raman intensity option implemented in CRYSTAL14 based on the coupled perturbed Hartree-Fock/Kohn-Sham (CPHF/KS) method. Since the output is restricted to intensities, it does not give access to the signs of the tensor elements. In order to obtain this property, we calculated the first-order dielectric tensor using the CPHF/KS method for the equilibrium crystal configuration and for the crystal with atomic displacement according to the vibrational movement for the individual phonon modes which was taken from the frequency calculation output. The comparison of the values in Tabs. \ref{tab:tensor_elements_Ag} and \ref{tab:tensor_elements_Bg} shows a good agreement between theory and experiment. This can also be seen from the plot of the modeled scattering intensity in Figs. \ref{fig:Ag_fits} and \ref{fig:Bg_fits}. For these, the Raman intensities were calculated setting the experimental conditions (temperature and excitation wavelength) in agreement to our experimental setup which is equivalent to multiplying with the prefactor $C(\omega\sub{P})$ from equation \eqref{eq:prefactor}. In particular, the general line shape is very well reproduced for most phonon modes, with distinct exceptions particularly for the $A_g^{(1)}$ and $A_g^{(8)}$ mode. Further, the intensity of the low energy modes $A_g^{(1)}$, $A_g^{(2)}$, $B_g^{(1)}$ and $B_g^{(2)}$ is strongly overestimated. Except for some deviations in the magnitude of individual Raman tensor elements, the agreement for the other phonon modes is very good.

In order to compare our results to the absolute intensities reported by Dohy \etal\ \cite{GaO_Raman_IR} for polarization directions parallel to principal axes of the indicatrix, the Raman tensor elements must be squared and multiplied with $C(\omega\sub{P})$ from equation. \eqref{eq:prefactor}. When doing so, almost identical values with only minor deviations to these results \cite{GaO_Raman_IR} are obtained. However, for the results to agree one has to assume that Dohy \etal\ confused the $x$- and $y$-axis defined by them similar to our definition. That means that their data for ``XX'' polarization corresponds to a polarization parallel to the $y$-axis of our dielectric coordinate system, which is close to the $[100]$-direction.

The remarkably good agreement between the plainly theoretical data combining ab-initio calculations and the model for anisotropic crystals \cite{RTFAC} opens an additional way to identify phonon modes beyond the simple assignment based on selection rules which can be ambiguous under certain conditions. By the comparison between the polarization dependencies from theory and experiment, not only the phonon symmetry, but the actual vibrational mode can be assigned. Usually, this was done solely based on the spectral position which might yield erroneous results. In order to make use of the polarization dependence of the Raman intensity for any crystallographic orientation, it renders necessary to not only investigate Raman intensities, but the actual Raman tensor elements, particularly including their sign. The full Raman tensor also in general allow to determine the crystallographic orientation of a given sample from the relative Raman intensities of the individual phonon modes. However, the latter task requires a high accuracy of the Raman tensor elements which cannot (yet) be provided by ab-initio calculations as shown by our results. For this problem, experimental Raman tensor elements, as determined here for \gao , are necessary. Nevertheless, due to the good agreement, it seems reasonable for example to assign the sign of the Raman tensor element $c$ of the $A_g$ phonon modes based on the ab-initio calculations.

\section{Conclusions}

We successfully studied the Raman spectrum of \gao\ with particular focus on the Raman intensity. We modeled the dependency of the intensity on the scattering configuration for most phonon modes of \gao , successfully applying the model for anisotropic crystals. From that, we obtained the experimental Raman tensor elements for these modes and found a good agreement with results from ab-initio calculations. The experimental accessibility of the Raman tensor element signs and their impact on the actual scattering intensities, which can be well-modeled, strongly suggest not only to investigate the absolute Raman intensities, but the tensor elements itself by both experiment and theoretical calculations.

\begin{acknowledgments}
We gratefully acknowledge Zbigniew Galazka from IKZ Berlin for supplying the (100)-oriented \gao -crystal and thank Daniel Splith for providing the software used to implement the modeling of the polarization-dependent intensities. Our Raman setup has been funded by Deutsche Forschungsgemeinschaft within Sonderforschungsbereich 762 ``Functionality of Oxide Interfaces''.
\end{acknowledgments}

\end{document}